\documentclass[twoside]{article}
\usepackage{fleqn,espcrc2}
\usepackage{graphicx}
\usepackage{amsmath}

\def\bea{\begin{eqnarray}}
\def\eea{\end{eqnarray}}

\newcommand{\as}{\alpha_{\scriptscriptstyle S}}
\newcommand{\Tau}{\tau}
\newcommand{\mtau}{m_\tau}
\newcommand{\mZ}{m_Z}
\newcommand{\asTau}{\as(\mtau^2)}
\newcommand{\asZtau}{\as^{(\tau)}(\mZ^2)}
\newcommand{\asZZ}{\as^{(Z)}(\mZ^2)}
\newcommand{\BR}{\cal{B}}
\newcommand{\nut}{\nu_\tau}
\newcommand{\nub}{\overline{\nu}}
\newcommand{\nueb}{\nub_{\text{e}}}
\newcommand{\Kbar    }{\kern 0.2em\overline{\kern -0.2em K}{}}
\newcommand{\Kb      }{\Kbar}
\newcommand\Rtau{R_\tau}
\newcommand\RtauV{R_{\tau,V}}
\newcommand\RtauA{R_{\tau,A}}
\newcommand\RtauS{R_{\tau,S}}
\newcommand\RtauVA{R_{\tau,V/A}}

\newcommand{\ointl}{\oint\limits}
\newcommand{\intl}{\int\limits}
\newcommand{\e}{\varepsilon}
\newcommand{\hm}{\hspace{-0.1cm}}

\title{ \boldmath Precise $\as$ from $\Tau$ Decays }

\author{ B. Malaescu\address{Laboratoire de l'Acc{\'e}l{\'e}rateur Lin{\'e}aire,
         IN2P3/CNRS et Universit\'e Paris-Sud 11 (UMR 8607),\\
         F--91405, Orsay Cedex, France} \thanks{The author would like to thank M. Davier, S. Descotes-Genon, A. H\"ocker, and Z. Zhang for their fruitful collaboration.
This work was supported in part by the EU Contract No.
MRTN-CT-2006-035482, \lq\lq FLAVIAnet''.}}

\begin{document}

\begin{abstract}
An updated measurement of $\asTau$ from ALEPH $\tau$ hadronic spectral functions
is presented.
We report a study of the perturbative prediction(s) showing that the fixed-order
perturbation theory manifests convergence or principle problems not presented 
in the contour-improved calculation.
Potential systematic effects from quark-hadron duality violations are estimated
to be within the quoted systematic errors.
The fit result is $\asTau = 0.344 \pm 0.005 \pm 0.007$, where the first error is
experimental and the second theoretical.
After evolution, the $\as(\mZ^2)$ determined from $\tau$ data is one of the most precise
to date, in agreement with the corresponding $N^3LO$ value derived from 
Z decays.
\end{abstract}

\maketitle


\section{INTRODUCTION}
The hadronic decays of the $\Tau$ lepton provide a clean laboratory to
perform precise studies of QCD.
Invariant mass distributions obtained from long distance hadron data 
allow one to compute the spectral functions, which permit the study of short 
distance quark interactions.
In particular, these spectral functions can be exploited to precisely determine
the strong coupling constant at the $\tau$-mass scale, $\asTau$ . 
Most of the present analysis is described in detail in ref~\cite{Davier:2008sk}.
Some new studies on the perturbative methods are described in Sec.~\ref{sec:thPred}.

\section{TAU HADRONIC DATA AND SPECTRAL FUNCTIONS}
The nonstrange vector (axial-vector) spectral functions $v_1(a_1)$, for a
spin 1 hadronic system, are obtained from the squared hadronic mass 
distribution, normalised to the hadronic branching fraction (with the hadronic observable(s)
$R_{\tau,V/A} = \frac{\BR_{\tau\to V^-/A^-(\gamma)\nut}}{\BR_{\tau\to {\text{e}}^- \nueb \nut}}$),
and divided by a factor exhibiting kinematics and spin characteristics
\begin{equation}
\label{eq:sf}
\begin{split}
   v_1(s)/a_1(s) 
   \propto 
              \frac{d N_{V/A}}{N_{V/A}\,ds}\,
              \frac{\BR_{\tau\to V^-/A^-(\gamma)\nut}}
                   {\BR_{\tau\to {\text{e}}^- \nueb \nut}} \\
              \left[ \left(1-\frac{s}{\mtau^2}\right)^{\!\!2}\,
                     \left(1+\frac{2s}{\mtau^2}\right)
              \right]^{-1}\hspace{-0.3cm}\:.
\end{split}
\end{equation}
The basis for comparing a theoretical description of strong interaction 
with hadronic data is provided by the optical theorem, which
relates the imaginary part of the polarisation functions on 
the branch cut along the real axis, to the spectral functions:
$2\pi \cdot Im\Pi^1_{V/A}(s) = v_1/a_1(s)$.

The determinations of $\Rtau$ from measured leptonic 
branching ratios, or only from the electronic one assuming universality, 
are in very good agreement, yielding
$ \Rtau = ({1-\BR_{\text{e}}-\BR_\mu})/{\BR_{\text{e}} } = {1}/{\BR_{\text{e}}^{\rm uni}}-1.9726 = 3.640 \pm 0.010\:. $
One can identify in $\Rtau$ a component with net strangeness and two nonstrange 
vector(V) and axial-vector(A) components.
Including the latest results from BABAR and Belle the value of the strange
component is $\RtauS = 0.1615 \pm 0.0040\:.$
The separation of the V and A components for final states with only pions 
is done using G-parity.
For the $K \Kb \pi$ mode, we assume CVC and use new results 
from the BABAR Collaboration~\cite{Aubert:2007ym}.
Finally, we get the components:
$ \RtauV	= 1.783 \pm 0.011 \pm 0.002$ and
$ \RtauA	= 1.695 \pm 0.011 \pm 0.002\:,$
where the first errors are experimental and the second due to the $V/A$ 
separation.

\section{THEORETICAL PREDICTION OF $\Rtau$}
\label{sec:thPred}
The nonstrange ratio $\RtauVA$ can be written as an integral of the 
spectral functions  over the invariant mass-squared $s$ of the final state hadrons 
\begin{equation}
\label{eq:rtauth1}
\begin{split}
& \RtauVA(s_0) \propto 
      \intl_0^{s_0}
		\frac{ds}{s_0}
      \left[ {\rm Im}\Pi^{(0)}_{V/A}(s+i\e) \right. \\
     & \left. \,+\, \left(1+2\frac{s}{s_0}\right){\rm Im}\Pi^{(1)}_{V/A}(s+i\e) \right]
       \left(1-\frac{s}{s_0}\right)^{\!\!2}. 
\end{split}
\end{equation}
The two point correlator can not be predicted by QCD in this region of the real 
axis. 
However, using Cauchy's theorem, one can relate this expression to an integral on
a circle in the complex plane.
Then, the OPE yields
 $\RtauVA \propto 1 + \delta^{(0)}
   + \delta^\prime_{\rm EW} + \delta^{(2,m_q)}_{ud,V/A}
   + \sum_{D=4,6,\dots}\delta_{ud,V/A}^{(D)} \:,$ 
with a massless perturbative contribution, a non-logarithmic electroweak
correction, the dimension two perturbative quark-mass contribution and higher 
dimension nonperturbative condensates contributions respectively.
The perturbative part reads 
$\delta^{(0)} = \sum_{n=1}^\infty \tilde{K}_n(\xi) A^{(n)}(\as) \:,$
with the functions 
\begin{equation}
\label{eq:an}
\begin{split}
  & A^{(n)}(\as) = \frac{1}{2\pi i}\hm\ointl_{|s|=s_0}\hm\hm
                  \frac{ds}{s} \cdot 
                  \left( a_s(-\xi s) \right)^n \\
                  & \cdot \left( 1 - 2\frac{s}{s_0} +          
                  2\left(\frac{s}{s_0}\right)^3 -      
                  \left(\frac{s}{s_0}\right)^4  \right)\:,
\end{split}
\end{equation}
where $a_s\hm\equiv\hm{\alpha_s}/{\pi}$ and $\xi$ is a scale factor.
A breakthrough was made recently~\cite{Baikov:2008jh}, so that the pertubative  
coefficients are now known up to $\tilde{K}_4$~(see~\cite{Davier:2008sk} for the numerical
values of the $\tilde{K}_n(\xi)$ coefficients).

\subsection{Perturbative methods}
The perturbative contribution to $\Rtau$ provides the main source of sensitivity 
to $\alpha_s(s_0)$.
The value of the strong coupling in the complex plane can be computed assuming
the validity of the renormalisation group equation~(RGE) outside the real axis,
and using a Taylor series of $\eta \equiv ln(s/s_0)$.

In the fixed order perturbation theory~(FOPT), at each integration step, the
Taylor expansion is made around the physical value $\as(s_0)$.
This can cause important problems as the absolute value of $\eta$ 
gets large and the convergence speed of the series is reduced~\cite{Davier:2008sk}.
In addition, a cut at a fixed order in $\as(s_0)$ is applied on the Taylor 
series and on the integration result in FOPT. 
Therefore, important known higher order terms are neglected, yielding additional
systematic uncertainties.

A better suited method is CIPT which, at each integration step, computes $\as(s)$
using the value found at the previous step.
In this approach the Taylor expansion is always used for small absolute values
of its parameter, hence excellent convergence properties.

One can analytically prove that the FOPT result can also be obtained by making
small steps, with a fixed order cut of the result at each step.
However, in that case the effective RGE is modified at every single step.
This is another way to see the problems of the FOPT method, which exist for a
transformation on the circle in the complex plane, as well as for a scale 
transformation on the real axis.

At the order $\beta_0$ one can analytically compute the solution of the RGE, 
\begin{eqnarray}
\label{eq:asan1}
  &&a_s^{an}(s) = \frac{a_s(s_0)}{1 + \beta_0 \cdot a_s(s_0) \cdot ln\left( \frac{s}{s_0}\right)}\\ 
\label{eq:asan2}
  &&\approx a_s(s_0) - \beta_0 \cdot \left(a_s(s_0)\right)^2 \cdot ln\left( \frac{s}{s_0}\right) \:.
\end{eqnarray}
On the real axis, as well as on the circle in the complex plane, CIPT reproduces
the analytical solution(\ref{eq:asan1}), whereas FOPT its approximation(\ref{eq:asan2}).

At higher orders, one can analytically compute the integral of the inverse 
beta function, yielding at the order $\beta_1$
\begin{equation}
\label{eq:intbm1}
\begin{split}
 \intl_{a_s(s_0)}^{a_s(s)}\hm\hm\frac{da}{\beta\left( a \right)} 
  =\hm\left[ \frac{1}{a \beta_0} + \frac{\beta_1}{\beta_0^2} 
   ln\hm\left( \frac{a}{\beta_0 + a\beta_1} \right)\hm\right]_{a_s(s_0)}^{a_s(s)}\hm\hm\:,
\end{split}
\end{equation}
and then relate it to $ln(s/s_0)$ using the RGE. 
Solving the resulting equation one gets a solution that we will call $\as^{an}(s)$.
It is also instructive to compare the results of the two methods with a solution
proposed for example in \cite{Yao:2006px}, consisting of an expansion in inverse
powers of logarithms
\begin{equation}
\label{eq:asPDG}
\begin{split}
 a_s^{PDG}(s)=&\frac{1}{\beta_0\cdot ln\hm\left( \frac{s}{\Lambda^2} \right)} 
      \hm\left[1-\frac{\beta_1}{\beta_0^2}\hm\cdot\hm
      \frac{ln\hm\left( ln\hm\left( \frac{s}{\Lambda^2} \right) \right)}{ln\hm\left( \frac{s}{\Lambda^2}\right)}
      \right.\\
      &\left. + O\left( \frac{1}{ln^2\hm\left( \frac{s}{\Lambda^2}\right)} \right)
      \right] \:.
\end{split}
\end{equation}

\vspace{-0.7cm}
\begin{figure}[htb]
\includegraphics[width=1.\columnwidth]{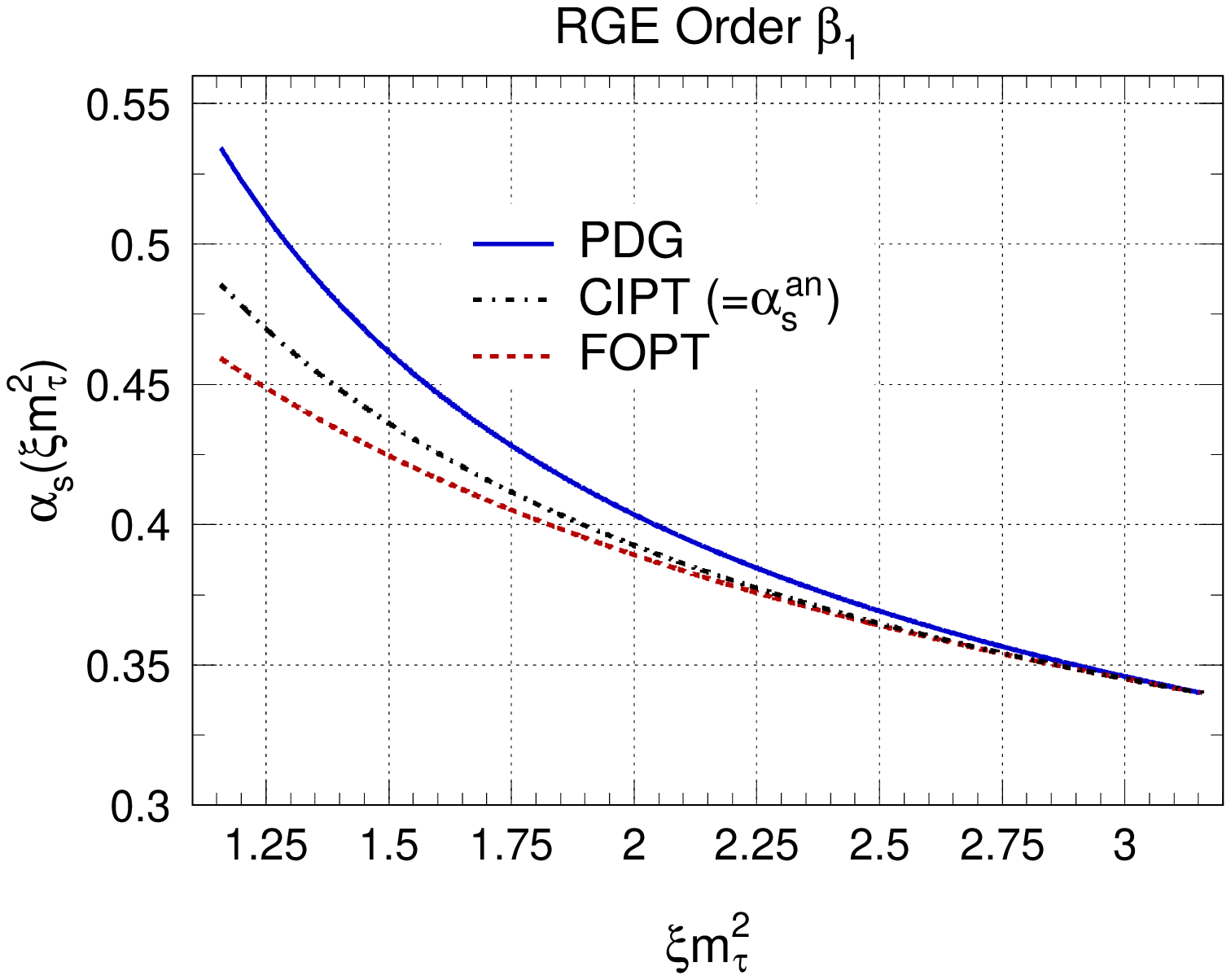}
\vspace{-1.3cm}
\includegraphics[width=1.\columnwidth]{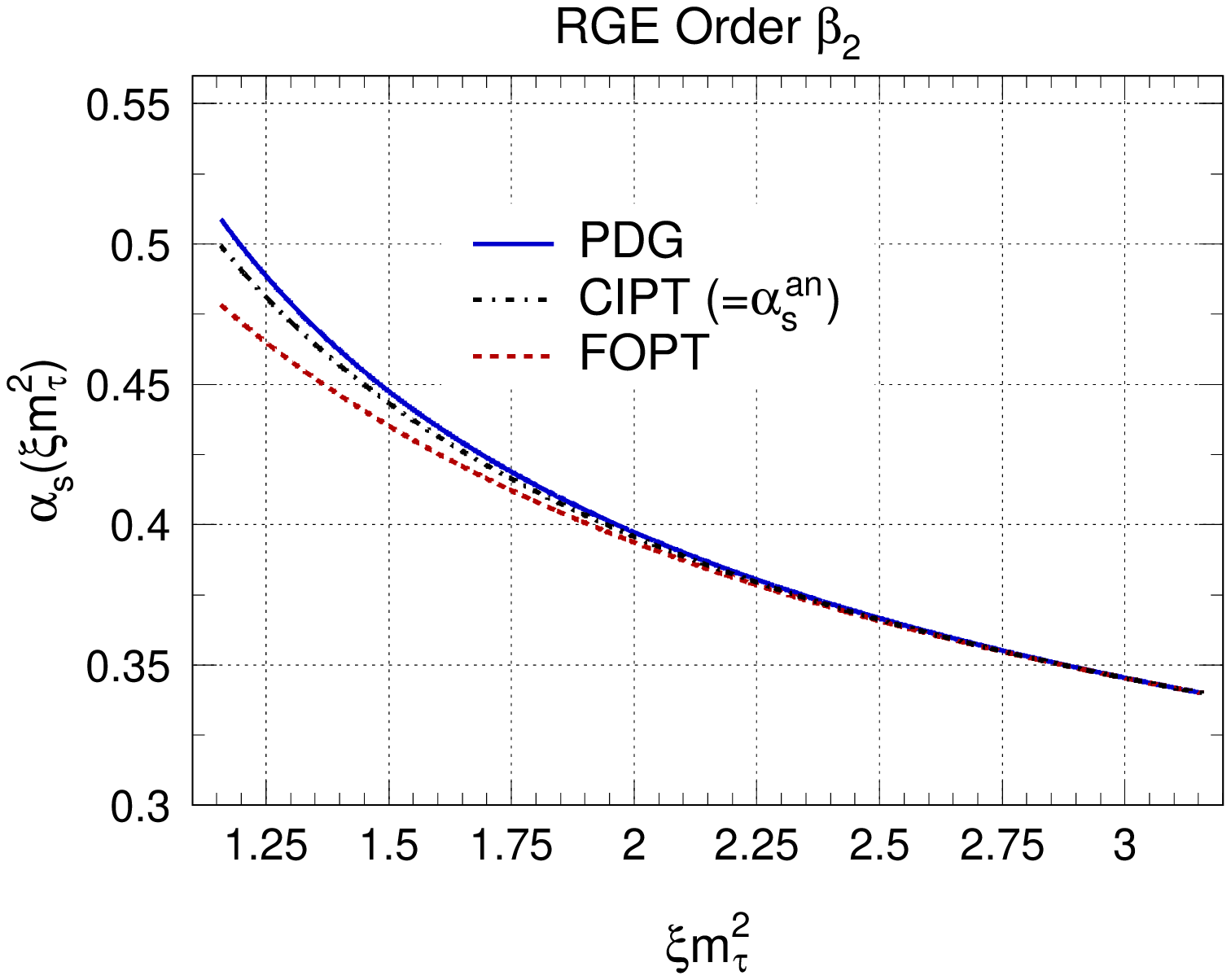}
\vspace{-0.5cm}
\caption{The evolution of $\as(s)$ to lower scales $\xi\mtau^2$ using 
         CIPT, FOPT, the analytical solution of the RGE ($\as^{an}(s)$) 
         and the expansion in inverse powers of logarithms ($\as^{PDG}(s)$).
         These solutions were computed up to the order $\beta_1$ of the RGE (top),
         and $\beta_2$ (bottom).}
\label{fig:evrealaxis}
\vspace{-0.6cm}
\end{figure}
In Fig.~\ref{fig:evrealaxis} we show a comparison of scale transformations performed
with CIPT, FOPT and the two other RGE solutions described above.
CIPT yields to the same result as the analytical solution, and the expansion in
inverse powers of logarithms is closer to it than to FOPT.
We should emphasize the fact that, on the contrary to CIPT, the FOPT solution
is problematic even on the real axis, generally not satisfying to the RGE.

For numerical applications we have used geometric growth estimations for the first unknown 
coefficients $\beta_4$, $K_5$ and $K_6$.
We have tested that CIPT is less sensitive to changes of these coefficients,
and it also exhibits a smaller scale dependence than FOPT.
Numerically, the difference of the perturbative contributions computed with the
two methods is about $15\%$.
In fact this difference could have been much larger if not for the properties
of the kernel in the integral (\ref{eq:an}) which has small absolute values in 
the region where the $\as(s)$ predictions of the two methods are rather 
different~\cite{Davier:2008sk}.

In ref.~\cite{Davier:2008sk}, we analysed further problems with the
perturbative series for FOPT when the integration over the circular
contour is performed.
CIPT behaves better than FOPT and is to be preferred.
The difference between the results obtained with the two approaches is not to
be interpreted as a systematic theoretical error, but rather like a problem
of FOPT~\cite{Davier:2008sk}.\footnote{See however a recent study with different
conclusions~\cite{Beneke:2008ad}.}

\subsection{Quark-hadron duality violation}
It is known that OPE describes only part of the nonperturbative effects.
In order to estimate the impact of the missing contributions, we test two models
based on resonances and on instantons. 
We add their contributions to the theoretical prediction, choosing parameters 
that provide a good fit to the V+A spectral function near the $\tau$ mass.
For these models, we find corrections situated within our systematic 
uncertainties~\cite{Davier:2008sk}.

\section{COMBINED FIT}
In order to obtain additional experimental information, we use spectral
moments defined as
\begin{equation}
   R_{\tau,V/A}^{kl} =
       \intl_0^{\mtau^2} ds\,\left(1-\frac{s}{\mtau^2}\right)^{\!\!k}\!
                              \left(\frac{s}{\mtau^2}\right)^{\!\!l}
       \frac{dR_{\tau,V/A}}{ds}\:.
\end{equation}
They allow one to better exploit the shape of the spectral functions and
they suppress the region where OPE fails.
The corresponding theoretical prediction is very similar to the one for $\Rtau \:,$ 
with consequent perturbative and nonperturbative contributions.
Due to strong correlations, we use only $R_\tau$~($k=0$ and $l=0$) and 
four additional moments~($k=1$ and $l=0,1,2,3$) to simultaneously fit 
$\asTau$ and the leading D = 4, 6, 8 nonperturbative contributions.

\begin{figure}[htb]
\includegraphics[width=1.\columnwidth]{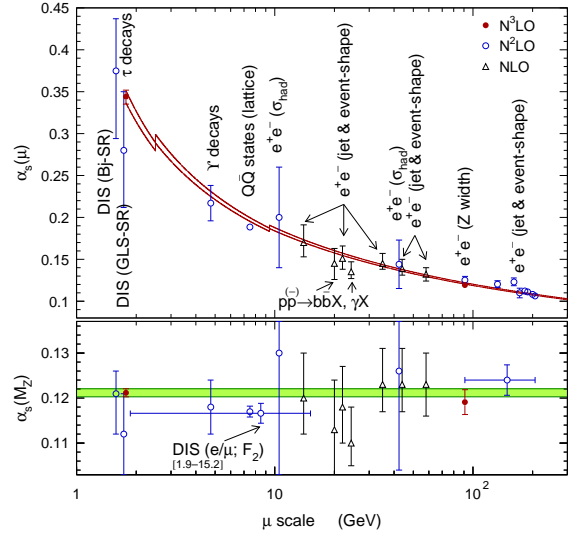}
\vspace{-1.5cm}
\caption{Top: The evolution of $\asTau$ to higher scales $\mu$ using 
         the four-loop RGE and the three-loop matching conditions applied at the
         heavy quark-pair thresholds (hence the discontinuities at $2\overline m_c$ 
         and $2\overline m_b$). The evolution is compared with independent 
         measurements covering $\mu$ 
         scales that vary over more than two orders magnitude. Bottom: The corresponding 
         $\as$ values evolved to $\mZ$. The shaded band displays the $\tau$ decay 
         result within errors.}
\label{fig:evolution}
\vspace{-0.3cm}
\end{figure}

In spite of the fact that the nonperturbative contributions fitted for the V and A
spectral functions have opposite signs and they are much larger 
than those from V+A, we find an excellent agreement between the values found for 
$\asTau$ from the three fits.
The result of the fit to the V+A spectral moments reads
\begin{equation}
\label{astau-res}
   \asTau = 0.344 \pm 0.005 \pm 0.007~,
\end{equation}
where the first error is experimental and the second is theoretical.
When evolving this value to the Z scale~\cite{Davier:2008sk}(see Fig. \ref{fig:evolution}) one
gets
\begin{equation}
\label{eq:asrez_mz}
\asZtau = 0.1212 \pm 0.0005 \pm 0.0008 \pm 0.0005\:,
\end{equation}
where the first two errors are propagated from~(\ref{astau-res}), and the last one 
summarises uncertainties in the evolution.
The consistency between this result and the value found by a global fit to electroweak
data at the Z-mass scale~\cite{Davier:2008sk}, $\asZtau-\asZZ = 0.0021 \pm 0.0029$, provides 
the most powerful present test of the evolution of the strong interaction coupling, over 
a range of $s$ spanning more than three orders of magnitude.

\section{CONCLUSIONS}
Motivated by some new results both on theoretical and experimental grounds, we have 
revisited the determination of $\asTau$ from the ALEPH $\tau$ spectral functions.
We have re-examined two common numerical methods: we have identified specific 
consistency problems of FOPT, which do not exist in CIPT.
The $\tau$ measurement of $\alpha_s$ evolved to the Z scale is found to be in excellent 
agreement with the direct determination from Z decays. 
Both results are the only ones at $N^3LO$ order so far, confirming the running of $\as$
between 1.8 and 91 GeV, as predicted by QCD, with an unprecedented precision of $2.4\%$.

\end{document}